\documentclass[prl,twocolumn]{revtex4}
\usepackage{color}
\usepackage{graphicx}
\usepackage{amsmath}
\begin{document}

\title{Gate-dependent tunneling-induced level shifts observed in carbon nanotube quantum dots}

\author{J.~V. Holm, H.~I.~J\o rgensen, K.~Grove-Rasmussen, J.~Paaske, K. Flensberg, P.~E.~Lindelof}
\affiliation{Niels Bohr Institute \& Nano-Science Center,
  University of Copenhagen, DK-2100 Copenhagen, Denmark}
\date{\today}

\begin{abstract}
We have studied electron transport in clean single-walled carbon
nanotube quantum dots. Because of the large number of Coulomb
blockade diamonds simultaneously showing both shell structure and
Kondo effect, we are able to perform a detailed analysis of
tunneling renormalization effects. Thus determining the environment induced level shifts of this artificial atom. In shells where only one of the
two orbitals is coupled strongly, we observe a marked asymmetric gate-dependence of the
inelastic cotunneling lines together with a systematic gate
dependence of the size (and shape) of the Coulomb diamonds. These
effects are all given a simple explanation in terms of second-order
perturbation theory in the tunnel coupling.
\end{abstract}

\maketitle

A solid-state quantum dot~(QD) is often referred to as an artificial
atom\cite{Kouvenhoven,Kouven2}, and the energy levels of this
artificial atom can be measured using bias spectroscopy in the
Coulomb blockade regime\cite{Tans,Bockrath,Cobden,Herrero}. In this
regime, the couplings to the metallic leads are usually regarded as
small perturbations, giving rise to a finite conductance without
otherwise affecting the internal structure and energy levels of the
QD. Deviations from the perfect shell structure in, for example,
carbon nanotubes are usually attributed to imperfections of the
single-walled carbon nanotube~(SWCNT). Consequences of stronger
tunnel coupling to the metallic leads have been studied in the
context of broadening of energy levels\cite{Liang} and Kondo effects \cite{Nygaard,Babic,Paaske}.

For moderate coupling strengths a certain degree of level shifts and
broadening is expected to arise from the hybridization with the
leads. Nevertheless, these effects are difficult to distinguish and
therefore rarely discussed in the literature. So far, the only case
where such renormalization effects have been shown to have important
observable consequences is for QD spin-valve systems, where either
non-collinear magnetization gives rise to off-diagonal
renormalization of the spin structure on the QD\cite{braigbrouwer05}
or spin-dependent tunneling gives rise to a gate-dependent
exchange-field on the QD~\cite{Martinek05,Hauptmann07}. In a similar
manner, one might expect a difference in couplings to the two
orbitals (or subbands) in the SWCNT to give rise to distinct
observable renormalization effects in a SWCNT.

In this Letter, we report on measurements on several SWCNT-devices
showing regular four-electron shell structure. The best device shows
a total of 60 regular four-electron shells, allowing for a unique
quantitative and statistical analysis of renormalization effects. In
particular, we study the cases where the four-electron shells
exhibit Kondo effect in only one of the diamonds in the quartet,
indicating that one of the (almost) degenerate orbitals has a
stronger coupling to the leads than the other. Such behavior is seen
in 28 four-electron shells. Based on this analysis we find that the
relative magnitudes of the tunnel coupling of orbitals one and two
determine three distinctly different types of quartets.
\begin{figure}[t]
\includegraphics[width=\columnwidth]{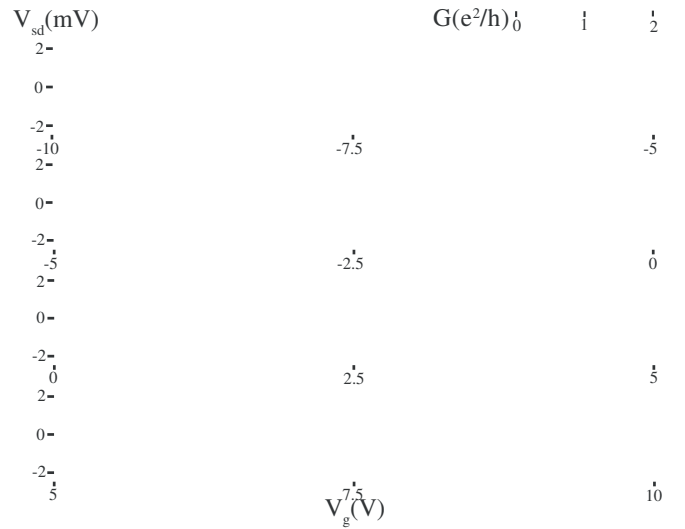}
\caption{\noindent \label{fig:fig1}Bias spectroscopy plot of a SWCNT
QD for $-10\mbox{V}<V_{g}<10\mbox{V}$. Coulomb diamonds are seen for
almost every added electron (285), and 88 odd-occupancy Coulomb blockade
diamonds exhibit a zero-bias Kondo resonance.}
\end{figure}

A difference in tunnel coupling for the two orbitals gives rise to
two features: 1) the inelastic cotunneling threshold given by the orbital splitting develops a marked gate dependence,
and 2) the addition energies of the individual Coulomb diamonds
within a quartet deviate from the constant-interaction model \cite{Kouven2}, depending on the relative
magnitude of the tunneling couplings. Both features can be explained by
level shifts of the many-body states of the QD, calculated here to second-order in the tunnel coupling.
\begin{figure*}[t]
\includegraphics[width=1\textwidth]{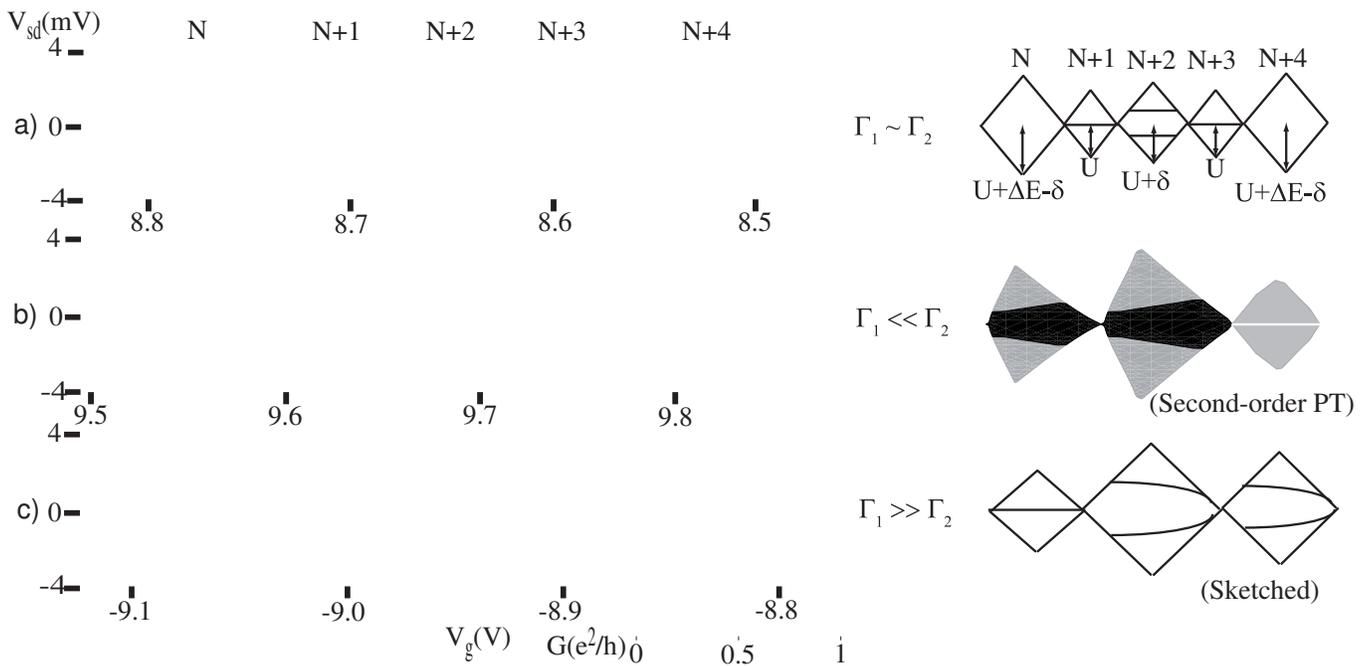}
\caption{\noindent \label{fig:fig2} Differential conductance as a
function of bias voltage, $V_{sd}$, and gate voltage, $V_{g}$
(left). Schematic of the four-electron shells (right). (a)
Four-electron shell filling for $\Gamma_{1}\sim\Gamma_{2}$, with
strong enough coupling to facilitate kondo ridges in the N+1 and N+3
diamonds, as shown in the schematic. A total of 17 four-electron
shells show this type of behavior across the full gate voltage
range. (b) Four-electron shell filling for
$\Gamma_{1}\ll\Gamma_{2}$. In the N+1 and N+2 diamonds, the
inelastic cotunneling threshold has clearly acquired a
gate-dependence. In total there are 21 four-electron shells showing
this type of behavior. Here the right panel shows the
cotunneling-thresholds and diamond edges as calculated within second-order PT (The white Kondo-ridge is drawn by hand). (c) Four-electron
shell filling for $\Gamma_{1}\gg\Gamma_{2}$. Gate-dependent
cotunneling thresholds are now observed in the N+2 and N+3 diamonds
instead. In total there are 7 four-electron shells showing this type
of behavior. 15 four-electron shells could not be uniquely
categorized. In the 28 shell-sequences where $\Gamma_1 \gg \Gamma_2$ or $\Gamma_1 \ll \Gamma_2$ the gate-dependent cotunneling threshold ridges inside the diamonds appear to form a U or V lying down with the openings facing the zero-bias Kondo ridge. Together the two U's (V's) and the Kondo ridge resemble a two-headed arrow, where the U's (V's) form the two heads and the Kondo ridge the shaft. This means that the tree center diamonds in the different domains have cotunneling and Kondo ridges, which can be schematically summarized as: $\Gamma_1 \sim \Gamma_2: \fbox{$-=-$},~\Gamma_1 \ll \Gamma_2: \fbox{$<<-$}$ and $\Gamma_1 \gg \Gamma_2: \fbox{$->>$}$.}
\end{figure*}

The devices were made on a highly n-doped $500\mu\mbox{m}$ thick
silicon substrate with a $500$nm thick SiO$_{2}$ top layer. Chemical
vapor deposition (CVD) was used to grow the SWCNTs from predefined
catalyst particle areas. The SWCNTs were contacted by thermal
evaporation of titanium/ aluminum/ titanium ($5$nm/ $40$nm/ $5$nm)
\footnote{The sample was cooled to 30mK, but due to a thin oxide
layer forming between the titanium and aluminum during evaporation,
no superconducting gap has been observed.}, and the electrodes were
separated by $300$nm and patterned by electron beam lithography. The
silicon substrate was used as a back-gate to change the
electrostatic potential of the SWCNT QD. The electron transport
measurements were carried out in a cryogenic insert system with a
base temperature of $300$mK. Two-terminal conductance was
measured using lock-in techniques with $13\mu\mbox{V}$ AC excitation
amplitude and a DC bias voltage ($V_{sd}$) applied to the source,
while the drain was grounded through a current amplifier.

Room temperature measurements of the conductance as a function of
back gate voltage ($V_{g}$) indicate that the SWCNT in the presented
device is metallic. In the full gate voltage span,
$-10\mbox{V}<V_{g}<10\mbox{V}$, four-electron shell-filling is
observed at low temperatures, and an overview of the regular behavior of
the device can be seen in Fig.~\ref{fig:fig1}. In total, 60 full
four-electron shells are observed across the measured gate voltage
range. Disregarding a few gate-switches, we have observed 285
consecutive electron additions. 60 four-electron shells can
be identified, the remaining 45 electrons are counted in incomplete
shells. We have extracted the Kondo temperatures from 88
equilibrium Kondo resonances from the width of the Kondo peak
($W_{FWHM}\sim 4 k_{B}T_{K}$)\cite{Micklitz06}. The Kondo
temperatures found by measuring the Kondo peak widths are in good
agreement with Kondo temperatures found for selected peaks, when
measuring the zero-bias conductance as a function of temperature,
and fitting to the universal scaling function\cite{Costi} (not
shown). We observe no systematic gate-dependence of the Kondo
temperatures, with an average of $\langle T_{K}\rangle\approx 1.2$K and standard deviation $0.5$K.

A more detailed bias spectroscopy plot of one of the four-electron
shells from Fig.~\ref{fig:fig1} can be seen in
Fig.~\ref{fig:fig2}~(a). The plot resembles previously reported
characteristics of a SWCNT QD in the Kondo
regime~\cite{Nygaard,Babic}. In diamonds N+1 and N+3, corresponding
to odd electron occupation, spin-1/2 Kondo ridges can be seen as
horizontal lines at zero bias voltage. In the N+1, N+2, and N+3
diamonds, the onset of inelastic cotunneling can be seen as
practically horizontal lines at
$V_{sd}\approx\pm0.7\mbox{mV}$~\cite{De Franceschi}. The cotunneling
ridge at $V_{sd}\approx+0.7\mbox{mV}$ in the N+2 diamond is
significantly Kondo enhanced~\cite{Paaske}. Equilibrium and
nonequilibrium Kondo features are observed in the full gate voltage
span. From the plot we can extract the characteristic energies of
the system as indicated in the schematic drawing on the right in
Fig.~\ref{fig:fig2}~(a)~\cite{Kouven2,Oreg}. We obtain a Coulomb
energy of $U=2.9\mbox{meV}$, orbital splitting
$\delta=0.7\mbox{meV}$ and single-particle level spacing $\Delta
E=3.5\mbox{meV}$.

In addition to this well-established four-electron shell behavior,
we have observed two distinctly different types of four-electron
shells, exemplified by the plots in Figs.~\ref{fig:fig2}~(b) and
(c). In Fig.~\ref{fig:fig2}~(b), there is a Kondo ridge in N+3, but
the expected Kondo ridge in the N+1 diamond is absent. In addition,
the cotunneling thresholds in the N+1, and N+2 diamonds increase
from left ($V_{sd}\approx\pm0.4
mV$) to right
($V_{sd}\approx\pm1.2mV$), showing almost identical gate-dependence.
In Fig.~\ref{fig:fig2}~(c), the situation is reversed: A zero-bias
Kondo ridge now only occurs in the N+1 diamond, and the cotunneling
thresholds in the N+2, and N+3 diamonds {\it decrease} from left
($V_{sd}\approx\pm0.4mV$) to right ($V_{sd}\approx0mV)$. The electron-hole symmetry (around the
middle of the N+2 diamond~\cite{Herrero}) which is observed in panel
(a) is clearly lifted in panels (b) and (c). Furthermore the main
features in panels (b) and (c) are each others mirror image.

As already mentioned, the sample shows a large number of
four-electron shells (60). 28 of them show only a single equilibrium
Kondo resonance and exhibit gate dependent inelastic cotunneling
lines inside the two adjacent Coulomb diamonds.

\begin{figure}
\includegraphics[width=1\columnwidth]{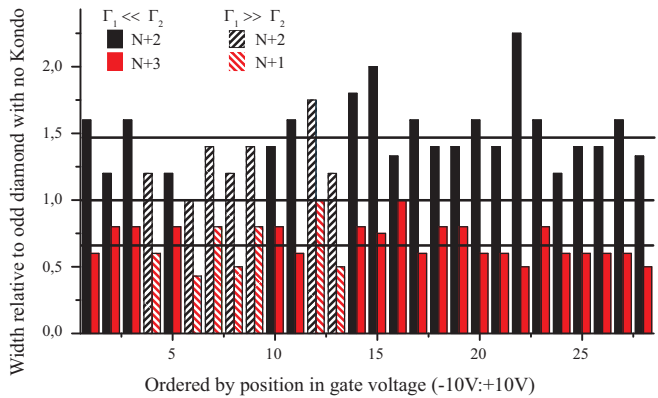}
\caption{\label{fig:fig3} (Color online) The ratio of the widths of the diamonds
(proportional to the addition energies) in the 28 unusual
four-electron shells to the odd-electron diamonds without a
Kondo ridge in the same four-electron shells. The solid
(cross-hathed) pairs is for $\Gamma_1\ll\Gamma_2$
($\Gamma_1\gg\Gamma_2$). The average width of the even-electron
diamonds (black) is 1.47, and the average width of the
odd-electron diamonds with a Kondo ridge (red (dark grey)) is
0.66, as indicated by the black lines.}
\end{figure}
Furthermore, there is a clear correlation between the size of the
odd-electron diamonds with, and without the zero-bias Kondo ridge,
respectively. This is demonstrated in Fig.~\ref{fig:fig3}, where
the widths of the $N+2$ diamonds (black) and the
odd-electron diamonds exhibiting a Kondo resonance
(red (dark grey)) are shown relative to the widths of the
odd-electron diamonds not showing a Kondo resonance.
For all 28 quartets of this type, the diamond showing a zero-bias
Kondo ridge is narrower than the diamond without the Kondo ridge,
independent of whether $\Gamma_1\ll\Gamma_2$ or $\Gamma_1\gg\Gamma_2$.
Observing this in 28 out of 60 shell-sequences, this behavior cannot
be dismissed as a measurement irregularity. We have seen similar
gate dependences in several other SWCNT devices with intermediate
coupling to the electrodes.

We shall now focus on the quartet mapped out in
Fig.~\ref{fig:fig2}~(b), where $\Gamma_1\ll\Gamma_2$.
The presence and absence of a zero-bias Kondo-ridge in the N+3, and
N+1 diamonds, respectively, already indicates that the two orbitals
must have different tunnel couplings, since the Kondo temperatures
must be respectively larger than or smaller than the base
temperature (300 mK). The Kondo temperatures are given by
$T^{(i)}_{K}\sim\sqrt{4\Gamma_{i}U}\exp(-\pi
U/(8\Gamma_{i})$~\cite{Hewson93}, in terms of the individual
tunnel-broadenings $\Gamma_i=\sum_{\alpha=s,d}\Gamma_{i,\alpha}$, with $\Gamma_{i,\alpha}=\pi\nu_{F}|t_{i,\alpha}|^{2}$,
determined by the density of states in the source, and drain
electrodes, $\nu_{F}$, and their respective tunnel couplings to the
dot, $t_{i,\alpha}$, assumed to be energy independent in the relevant energy range. The lack of particle-hole symmetry in this
quartet is thus consistent with the orbital symmetry being lifted by
a difference between $\Gamma_{1}$ and $\Gamma_{2}$. The $\Gamma_{i}$
are generally difficult to extract, but by fitting the even valley part
of the Coulomb peaks of Fig.~\ref{fig:fig2}~(b) to a Lorentzian (cf.
Ref.~\onlinecite{HIJ}), we estimate that
$\Gamma_{1}\approx0.08\mbox{meV}$ and
$\Gamma_{2}\approx0.46\mbox{meV}$. This yields
$T^{(1)}_{K}\sim7\mu\mbox{K}$ and $T^{(2)}_{K}\sim2\mbox{K}$,
consistent with the observation of a Kondo ridge only in the N+3
diamond at $T\sim 300$~mK.

The finite-bias cotunneling lines inside the N+1 and N+2 diamonds of
Fig.~\ref{fig:fig2}~(b) can be ascribed to an inelastic cotunneling
processes in which an electron traversing the dot from source to
drain has enough energy to excite a single electron from orbital one
to orbital two. The individual many-body states of the dot electrons
are renormalized by virtual charge-fluctuations, and, as we shall
argue below, these tunneling-induced level shifts can indeed give
rise to a gate-dependent cotunneling threshold.

\begin{figure}[t]
\includegraphics[width=\columnwidth]{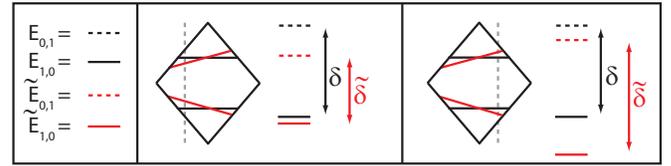}
\caption{\noindent (Color online) Schematic drawing illustrating the difference in
renormalization of ${\tilde E}_{1,0}$ (red (dark grey) solid lines) and ${\tilde
E}_{0,1}$ (red (dark gray) dashed lines), at opposite ends of the $N+1$-diamond.
Assuming $\Gamma_{1}\ll\Gamma_{2}$, the (0,1)-level is strongly
perturbed by virtual emptying (left), whereas the (1,0)-level is
most strongly perturbed by virtual double-occupancy (right).
\label{fig:fig4}}
\end{figure}

Within a single four-electron shell, the many-body states of the dot
electrons can be enumerated by the number of electrons occupying the
two highest lying single-particle orbitals. In terms of bare
level-position, $\varepsilon_{d}$, intra-dot Coulomb repulsion, $U$,
and orbital splitting, $\delta$, the corresponding energies are
given by $E_{i,j}=(i+j)\varepsilon_d+j\delta+(i+j)(i+j-1)U/2$,
where $i(j)$ denotes the number of electrons in orbital $1(2)$.
The level shifts, $\delta E_{i,j}={\tilde E}_{i,j}-E_{i,j}$, are
determined within second-order many-body perturbation theory (PT) in the
tunnel coupling, by considering all possible fluctuations
experienced by a given charge configuration~\cite{Haldane}. For example, the charge
state $|1,0\rangle$ connects to $|0,0\rangle$ and to $|2,0\rangle$
and $|1,1\rangle$, respectively, by virtual tunneling {\it out}, and
{\it in} of an electron. This leads to the following shift in energy:
\begin{eqnarray}
\delta E_{1,0}\!\!&=&\!\!\!\sum_{\alpha=s,d}
\int_{-D}^{D}\!\frac{d\omega}{\pi}\,\mathrm{Re}\!\left[
\frac{\Gamma_{1\alpha}(1-f(\omega-\mu_{\alpha}))}
{E_{1,0}-(E_{0,0}+\omega)+i\Gamma}
\right.\label{Etilde}\\
&
&\left.\hspace*{-5mm}+\frac{\Gamma_{1\alpha}f(\omega-\mu_{\alpha})}
{E_{1,0}+\omega-E_{2,0}+i\Gamma}
+\frac{2\Gamma_{2\alpha}f(\omega-\mu_{\alpha})}
{E_{1,0}+\omega-E_{1,1}+i\Gamma}\phantom{\int_{-D}^{D}}
\hspace{-7mm}\right],\nonumber
\end{eqnarray}
where $2D$ is the conduction-electron bandwidth, $f(\omega)$ is the Fermi
function, and $\mu_{s,d}$ the chemical potentials of the source, and
drain electrodes. A level broadening of order $\Gamma\sim\Gamma_{1}+\Gamma_{2}$ has been included in the energy denominators and serves to cut off the otherwise logarithmically singular result. Notice that the last process has multiplicity two
from spin, and that only this process has the amplitude
$\Gamma_{2}$. The excited state is renormalized correspondingly and
$\delta E_{0,1}$ is given by Eq.~(\ref{Etilde}) with all orbital
indices reversed, i.e. $E_{i,j}\to E_{j,i}$ and
$\Gamma_{1\alpha}\leftrightarrow\Gamma_{2\alpha}$. This means that for
gate-voltages close to the 0/1 (left diamond corner), or 1/2 (right diamond corner), charge-degeneracy point,
respectively, it is the first, or the two last terms in
Eq.~(\ref{Etilde}) which dominate the renormalization, with small energy denominators. When
$\Gamma_1\ll\Gamma_2$, the $E_{0,1}$ energy level (orbital 2) will have a large
shift at the left corner of the diamond (0/1) and the $E_{1,0}$ energy level (orbital 1) will have a large shift at the right corner (1/2), as illustrated in Fig.~\ref{fig:fig4}.
This means that the cotunneling threshold increases from left to right when $\Gamma_1\ll\Gamma_2$, and decreases from left to right when $\Gamma_1\gg\Gamma_2$. To be more precise, the inelastic cotunneling threshold inside the N+1 diamond
is determined as the energy, $\tilde{\delta}$, solving the equation
$\tilde{\delta}=\tilde E_{1,0}-\tilde E_{0,1}$, with
$\mu_s - \mu_d=\pm\tilde{\delta}/e$ in the expressions for the $\tilde
E_{i,j}$. The solution, with parameters found above~\footnote{As a rough estimate we use $\delta$ and $U$ found from Fig.~\ref{fig:fig2}~(a)}, is plotted in
the right panel of Fig.~\ref{fig:fig2}~(b), with a reasonable
resemblance to the gate-dependence observed in the left panel.
Expanding near the middle of the $N+1$ diamond, we find that
$d\tilde{\delta}/d\varepsilon_d\approx
-4(\Gamma_{2}-\Gamma_{1})/(\pi U)$, for $\varepsilon_d=-U/2$.
Thus if $\Gamma_2>\Gamma_1$ the  cotunneling threshold clearly
increases with gate voltage ($eV_g\propto -\varepsilon_d$), i.e. in
the direction towards the stronger coupled $N+3$ diamond, as we
observe.

Calculating all the relevant level shifts, one can also determine
the changed gate voltage width, $\Delta V_g(n)$, of the $N+n$
diamond. We find that
$\Delta V_g(1)\approx U-(2\Gamma_1/\pi)\ln(U/\Gamma)$, $\Delta
V_g(2)\approx U+\delta$ and $\Delta V_g(3)\approx
U-(2\Gamma_2/\pi)\ln(U/\Gamma)$, strictly valid only in the regime
where $\delta\ll\Gamma\ll U$. A larger $\delta$ implies more correction terms,
but the main trend remains the same: a large $N+2$ diamond and a small $N+3$ diamond, as compared to the
$N+1$ diamond ($\Gamma_{1}\ll\Gamma_{2}$) in agreement with the
statistics presented in Fig.~\ref{fig:fig3}.

Finally, we have calculated the shifted Coulomb diamond
boundaries, using the second-order level shifts. For the found
parameters, this gives rise to the diamond boundaries seen in right
panel of Fig.~\ref{fig:fig2}~(b). The details of this figure are not
entirely reliable, and more work is needed in order to proceed
beyond second-order PT and to incorporate the nonequilibrium mixing of
ground, and excited states within each diamond. Nevertheless, this
simple analysis indicates that tunneling-induced level shifts may
give rise to skewed (i.e. non-parallel edges), rather than merely
shifted and rescaled Coulomb diamonds, which is in fact seen in
several of our experimental quartets, e.g. Fig.~\ref{fig:fig2}~(c).

In conclusion, we have fabricated gated SWCNT devices with
intermediate transparency contacts. Transport measurements have
revealed three distinct types of four-electron shells, two of which
show a marked gate-dependence of the inelastic cotunneling ridges,
together with a systematic pattern for the individual diamond-widths
within a shell. Calculating second-order level shifts for the
many-body states on the QD, we have explained all of these
observations in terms of a difference in tunnel coupling to the two
orbitals in the SWCNT.

We acknowledge support from EU projects ULTRA-1D, SECOQC, CARDEQ,
and the Danish Research Council.

\end{document}